# ОБЗОР МЕТОДОВ ПРИВЕДЕНИЯ БАЗИСА РЕШЕТОК И НЕКОТОРЫХ ИХ ПРИЛОЖЕНИЙ

**В. С. Усатюк**
**Научный руководитель: д.ф.-м.н., проф. О.В. Кузьмин**
**Братский государственный университет,**
**г. Братск, L@Lcrypto.com**

Широкий класс задач теории оптимизации, кодирования и криптографии решается методами геометрии чисел. Путем сведения их к проблеме приведения базиса решеток, к таковым относятся, в частности:

- задача комбинаторной оптимизации, входящая в «список Карпа» – поиск суммы подмножества (subset sum problem), криптоанализ системы шифрования Меркле-Хеллмана и ее производных, [1];
- задача о совместных диофантовых приближениях, решаемая на одном из этапов синтеза элементов управления, [2,3];
- задача оптимальной планировки памяти: решаемая планировщиком ОС, а так же компилятором, на стадии определения размера и стратегии обращения к сегменту данных, буферу в «куче». PE/COFF, ELF, Mach-0 секций исполняемых файлов в ОС семейства Windows, Linux, Mac OS X соответственно, [4];
- задача декодирования кодовых групп при передаче информации n-антеннами и приеме m-антеннами (MIMO), в частности применяемая в стандартах WIFI и WIMAX, [5,6];
- оценка экстремума (классов) функций многих переменных, для решеток эквивалентных положительным квадратичным формам, [7,8];
- задача синтеза и криптоанализа систем шифрования на основе задач теории решеток (Lattice based cryptography), [9].

**Определение 1:** Решетка - дискретная абелева подгруппа, заданная на множестве $R^m$. Решетку $L$ можно представить как множество целочисленных линейных комбинаций, $n$ линейно-независимых базисных векторов $\{\overline{b}_1,...,\overline{b}_n\} \subset R^m$ в $m$-мерном евклидовом пространстве: $L(b_1,...,b_n) = \{\sum_{i=1}^{n} x_i b_i : x_1,...,x_n \in Z^n\}$, где $m$ и $n$, размерность и ранг решетки соответственно.

**Определение 2:** Пусть дан базис решетки $B = \{\overline{b}_1,...,\overline{b}_n\} \in R^{m \times n}$, фундаментальным параллелипипедом заданым на этом базисе $P(B)$ называется множество: $P(B) = \left\{\sum_{i=1}^{n} x_i \cdot b_i : x_i \in [0,1)\right\}$.

**Определение 3:** Пусть дан базис решетки $B = \{\bar{b}_1, ..., \bar{b}_n\} \in R^{m \times n}$. Определитель решетки заданный этим базисом $\det(L(B))$ будем называть объем $n$-мерного фундаментального параллелипипеда образованного этим базисом $vol(P(B))$ : $\det(L(B)) = vol(P(B))$.

**Определение 4:** Решетки $L_1$, $L_2$, заданные базисами $B = \{\bar{b}_1, ..., \bar{b}_n\}$, $B' = \{\bar{b}'_1, ..., \bar{b}'_n\}$, конгруэнтны(эквивалентны друг другу), $L_1(B) \cong L_2(B')$, если определители этих решеток равны $\det(L_1) = \det(L_2)$, где $\det L = \left|\sqrt{\det(B^T B)}\right|$ или $\det L = |\det B|$ для полноразмерных(полноранговых) решеток.

Множество таких базисов $B_i$ может получено в результате умножения приведенного базиса решетки $L_i$ на целочисленные унимодулярные матрицы. Например, образованные базисами $B_1 = \{(1,1)^T, (4,0)^T\}$ и $B_2 = \{(8,4)^T, (9,5)^T\}$, решетки $L(B_1) \cong L(B_2)$, так как $|\det(B_1)| = |\det(B_2)| = abs\begin{pmatrix} 1 & 4 \\ 1 & 0 \end{pmatrix} = abs\begin{pmatrix} 8 & 9 \\ 4 & 5 \end{pmatrix} = 4$.

**Определение 5:** Пусть имеется радиус сферы заданный нормой $p$. Под $i$-м минимумом $\lambda_i^p(L)$ будем понимать наименьший радиус сферы, содержащий $i$-линейно независимых нетривиальных базисных векторов решетки. Длине кратчайшего вектора в решетке соответствует $\lambda_1^p(L)$.

Задача приближенного поиска кратчайшего базиса ($\gamma$-approximate shortest basis problem, $\text{S}BP_\gamma^p(n)$): Пусть дана $m$-мерная решетка $L$, ранга $n$, заданная базисом $B = \{\bar{b}_1, ..., \bar{b}_n\}$. Найти базис конгруэнтной решетки $B' = \{\bar{b}'_1 ..., \bar{b}'_n\}$: $\bar{b}'_1 = \gamma_1 \cdot \lambda_1^p(L), .., \bar{b}'_n = \gamma_n \cdot \lambda_n^p(L)$, где $\gamma_i \geq 1$.

**Определение 6:** $QR$-разложением базиса решетки называют представление матрицы $B = (\bar{b}_1, \bar{b}_2, ..., \bar{b}_n)$, размера $m \times n$ в виде произведения $Q$-унитарной и $R$-верхнетреугольной матриц:

$$B = (\bar{b}_1, \bar{b}_2, ..., \bar{b}_n) = Q^T \times R = (q_1, q_2, ..., q_n) \begin{pmatrix} r_{1,1} & r_{1,2} & \cdots & r_{1,n} \\ 0 & r_{2,2} & \cdots & r_{2,n} \\ \vdots & \ddots & \ddots & \vdots \\ 0 & \cdots & 0 & r_{n,n} \end{pmatrix}.$$

*QR*-разложение вычисляется методами: ортогонализации Грама-Шмидта, поворота Гивенса, отражения Хаусхолдера.

**Определение 7, [10]:** Базис решетки $L(B)$ образованный векторами $B = \{\overline{b}_1,...,\overline{b}_n\} \subset R^m$ называют приведенным по длинне (ослабл. Эрмиту), если для коэффициэнтов верхнетреугольной матрицы $R$, *QR*-разложения базиса решетки $B$ выполняется: $|r_{i,j}| \leq \frac{1}{2}|r_{i,i}|, 1 \leq i \leq j \leq n$.

**Определение 8, [11]:** Базис решетки $L(B)$ является с-приведенным базисом решетки, если он приведен по длине и для коэффициэнтов $R$, *QR*-разложения базиса решетки $B$: $r_{i-1,i-1}^2 \leq c \cdot r_{i,i}^2, 2 < i \leq n, c \geq \frac{4}{3}$.

**Определение 9, [12]:** Базис решетки $L(B)$ приведен по Ловасу, если он приведен по длине и для коэффициэнтов $R$ *QR*-разложения базиса решетки $B$ выполняется: $r_{i,i}^2 + r_{i-1,i}^2 \geq \delta r_{i-1,i-1}^2, 1 < i \leq n, \delta \in (0.25,1]$.

**Определение 10, [10, 13]:** Базис решетки $L(B)$ приведен по (Эрмиту)-Коркину-Золотареву(HKZ, KZ), если он приведен по длинне и для всех квадратных подматриц $R'$, размера $(n-i+1), 1 \leq i < n$, полученных из $R$ вычеркиванием $i$-первых строк и столбцов, первый вектор столбец является кратчайший в решетке $L'(R')$, $\overline{r}_1'^{(i)} = \lambda_1^p(L')$.

**Определение 11, [14]:** Базис решетки $L(B)$ приведен по Минковскому, если длины векторов базиса биективны соотвествующим минимумам в решетке, $SBP_{\gamma_1,\gamma_2,...,\gamma_m=1}^p(n)$.

На настоящий момент неизвестны полиномиальные алгоритмы приведения базиса по Минковскому, за исключением решеток размерности 2, предложенный Гауссом [15] и 3 предложенный Валле (с кубической сложностью) [16] и улучшенный Семаевым, снизившим сложность до квадратичной [17]. В работе [18] описан алгоритм для приведения базиса по Минковскому до 6 измерений. В работе [19] результат был улучшен до 7 мерных решеток. Для приведения базиса решетки по Ловасу применяется полиномиальный LLL-алгоритм предложенный Ленстра-Ленстра-Ловасом, **[**12]. Для приведения по Коркину-Золотареву применяется блочный метод Коркина-Золотарева (SE BKZ) предложенный Шнором-Охнером в работе [20].

*Литература*: